\documentstyle[preprint,aps,prb]{revtex}
\begin{document}
\title{Schr\"odinger Equation for Heavy Mesons Expanded in $1/m_Q$}
\author{Takayuki Matsuki
\thanks{E-mail: matsuki@tokyo-kasei.ac.jp or matsuki@ins.u-tokyo.ac.jp}}
\address{Tokyo Kasei University,
1-18-1 Kaga, Itabashi, Tokyo 173, JAPAN}
\maketitle
\begin{abstract}
Operating just once the naive Foldy-Wouthuysen-Tani transformation on
the Schr\"odinger equation for $Q\bar q$ bound states described by a
hamiltonian, we systematically develop a perturbation theory in
$1/m_Q$ which enables one to solve the Schr\"odinger equation to
obtain masses and wave functions of the bound states in any
order of $1/m_Q$. It is shown that positive energy projection
with respect to the heavy quark sector of a wave function is, at each order of
perturbation, proportional to the 0-th order solution. There appear also
negative components of the wave function except for the 0-th order,
which contribute also to higher order corrections to masses.
\end{abstract}
\section{Introduction}
Since the introduction of the ingenious notion of heavy quark
effective theory (HQET),\cite{HQET}
many physical quantities, especially regarding to
$B$ mesons, have been calculated. Furthermore formulation
has also been polished to incorporate higher orders.\cite{POL} However,
the way to incorporate higher order terms is somewhat insufficient
since only the operator product forms in higher orders are obtained but
their matrix elements are just parameters and must be determined so that the
physical quantities are fitted with experiments.\cite{REVIEW}
Another obscure point of HQET is that the residual momentum,
which is a difference between a
meson momentum and a heavy quark momentum, necessarily comes in
and we do not know how properly this quantity can be taken into
account as a perturbation.
On this point HQET
is different from QED in the sense that the next order perturbation terms
cannot be determined automatically from the lower orders.

Let us remind how HQET has been constructed to calculate physical quantities
regarding heavy mesons/baryons. When HQET was originally developed, lagrangian
formulation was adopted and light/heavy quarks are treated as
elementary fields. In this approach an appropriate transformation
only on heavy quarks is introduced
so that heavy quark fields are separated into positive/negative
components in energy. This picture was intuitively believed to realize
the situation that a positive component of
a heavy quark behaves like a static color source and a light
anti-quark with gluon clouds is relativistically hanging around it,
which owes most of the messy part of the strong interactions.
To stress even more this intuitive idea,
one assumes that a heavy quark carries most of the momentum of a heavy meson,
which yields a residual momentum.

Even though the above formulation has been well developed, this cannot
be applied straightforwardly to the Schr\"odinger equation for a
bound state of a $Q\bar q$ system.
It is the purpose of this paper to construct a perturbation
theory in $1/m_Q$ {\it a la} HQET for a hamiltonian system,
i.e., for a bound system which
is described by a hamiltonian with an appropriate potential.
Instead of particle fields, wave functions
of a $Q\bar q$ system {\it at rest} are the objects to deal with.
Then matrix elements of currents/operators between heavy mesons are
calculable at any order in principle
since meson wave functions are calculated as solutions to the Schr\"odinger
equation together with their masses as eigenvalues with corrections and
there appears no residual momentum of a heavy quark in this
formulation. In our formulation, at first the naive Foldy-
Wouthuysen-Tani (FWT)\cite{FWT} transformation is introduced and
instead of working only with positive
components of heavy quark fields there appear negative components as well
and hence the formulae to calculate those negative components are also
given at each order of perturbation. This formulation gives diffrent
results from the previous calculations\cite{Morii1} on the masses of
$D$, $D^*$, $B$ and $B^*$ because of negative components of heavy quark
sector which is briefly mentioned at the end of this letter.
\section{Hamiltonian}
\label{sec:hamil}
A projection operator which extracts a positive component of a heavy quark
plays an important role in HQET. In a much more precise statement this
projection operator is defined so that the equation for
the two upper components of a Dirac spinor (positive components)
does not mix with the two lower components (negative components)
in the lagrangian. Here before introducing a projection operator,
let us recall what the naive FWT transformation does.
This transformation, when it operates on
a free Dirac Schr\"odinger equation,
turns a kinematic hamiltonian, $\vec\alpha\cdot\vec p+m\beta$, into
$E \beta$ and leaves us only the two upper components of a free wave
function.
\[U_{FWT}\left( {\vec \alpha \cdot \vec p+m\beta } \right)
U^{-1}_{FWT}\;U_{FWT}\phi =E\;U_{FWT}\phi
\quad\to\quad E\;\beta \;U_{FWT}\phi = E\;U_{FWT}\phi,\]
i.e.,
\[E\left( {1-\beta } \right)\;U_{FWT}\phi =0,\]
which means that $(1+\beta)/2~U_{FWT}\phi=U_{FWT}\phi$, i.e.,
only the two upper components of $~U_{FWT}\phi$ survive.
Here
\begin{eqnarray}
  U_{FWT}\left( p \right)&=&\exp \left( {W\left( p \right)
  \,\vec \gamma \cdot \vec n} \right)=\cos W+\vec \gamma \cdot \vec
  n\,\sin W,\\
  \vec n&=&{{\vec p} \over p},\quad \tan W\left( p \right)={p \over {m+E}}.
\end{eqnarray}
As one can see, in this formulation a projection operator, $(1-\beta)/2$,
naturally comes in to exclude two lower components.
Based on this fact, introduction of the FWT transformation
may give us a good starting point for studying a $Q\bar q$
system as in fact will be seen later rather than trying to
develop a perturbation theory of the original hamiltonian
without any transformation.

The state of a bound state in this paper is defined by
\begin{equation}
   \left| \psi\right> = \int d^3 x \int d^3 y~\psi_{\alpha \beta}(x-y)~
  {q_\alpha^c~}^\dagger (x)~Q_\beta^\dagger (y) \left| 0 \right>,
\end{equation}
where $q^c$ is a charge conjugate field of a light quark $q$ XSand
its conjugate state by
$\left< \psi\right|=\left| \psi\right>^\dagger$
with $\left<0\right|\equiv\left|0\right>^\dagger$.
Given a hamiltonian density, $\cal{H}$, for a $Q\bar q$ system,
varying the quantity
\[ \left< \psi\right|\left({\cal H}-E\right)\left| \psi\right>, \]
in terms of $\psi_{\alpha \beta}(x-y)$, and setting it equal to zero,
then we have the matrix form of the Schr\"odinger equation
\[H\;\psi = (m_Q + \tilde E)\;\psi.\]
Here a bound state mass $E$ is broken into two parts, $m_Q$ and
the rest, $\tilde E$. A binding energy is given by $\tilde E-m_q$.
Operating the FWT transformation only on a heavy quark sector in
this equation at the
center of the mass system of a bound state, one can modify
the Schr\"odinger equation as,
\begin{equation}
   \left(H_{FWT}-m_Q\right)~\psi _{FWT}=\tilde E\;\psi _{FWT},\label{Schr}
\end{equation}
where
\begin{equation}
    H_{FWT}=U_{FWT}\left(p^\prime_Q\right)HU^{-1}_{FWT}\left(p_Q\right),
    \quad \psi _{FWT}=U_{FWT}\left( p_Q \right)\psi.
\end{equation}
Note that the argument of the FWT transformation operated on a hamiltonian
from left is different from the right-operated one,
since an outgoing momentum, ${\vec p_Q~}^\prime$, are different from incoming
one, $\vec p_Q$. Expressing model-dependent instantaneous
potential terms as $H^{int}$, a hamiltonian is given by
\begin{equation}
    H=\vec \alpha _q\cdot \vec p_q+m_q\beta _q+
    \vec \alpha _Q\cdot \vec p_Q+m_Q\beta _Q+H^{\rm int }.
\end{equation}
Next $H_{FWT}-m_Q$ can be expanded in $1/m_Q$, which is given by
\begin{equation}
    H_{FWT}-m_Q = H_{-1}+H_0+H_1+H_2+\cdots ,
\end{equation}
where
\begin{mathletters}
\label{hameqs}
\begin{eqnarray}
  H_{-1}&=&m_Q\left( {\beta _Q-1} \right),\label{ham_1}\\
  H_0&=&\vec \alpha _q\cdot \vec p+m_q\beta _q + H^{int}_{FWT0},\label{ham0}\\
  H_1&=&{1 \over {2\,m_Q}}{\vec p~}^2\beta _Q+H^{int}_{FWT1},\label{ham1}\\
  H_2&=& H^{int}_{FWT2}\label{ham2},\\
  \vdots\nonumber
\end{eqnarray}
\end{mathletters}
Here $H_i$ stands for the $i$-th order hamiltonian,
$H^{int}_{FWTi}$ for the $i$-th order of the interaction terms and
since a bound state is at rest,
\[
    \vec p=\vec p_q=-\vec p_Q,\quad
    {\vec p}~^\prime ={\vec p_q}~^\prime=-{\vec p_Q}~^\prime,\quad
    \vec q={\vec p}~^\prime - \vec p,
\]
are defined, where primed quantities are final momenta.
\section{Perturbation}
\label{sec:pert}
Using the hamiltonian obtained in the last section, we show in this
section that the Schr\"odinger equation can be solved order by order in
$1/m_Q$ utilizing a projection operator. First we introduce projection
operators:
\begin{equation}
    \Lambda_\pm = {{1\pm \beta_Q}\over 2},
\end{equation}
which correspond to positive-/negative-energy projection operators
for a heavy quark sector at the rest frame of a bound state. These are
given by $(1\pm v{\kern-6pt /})/2$ in the moving frame of a bound state
with $v^\mu=P^\mu/E$ in HQET where $P^\mu$ is the
four-momentum of a bound state.
Then we expand the mass and wave function of a bound state in $1/m_Q$ as
\begin{eqnarray}
  \tilde E=E^k_0+E^k_1+E^k_2+\ldots,\\
  \psi_{FWT}=\psi^k_0+\psi^k_1+\psi^k_2+\ldots,
\end{eqnarray}
where $k$ stands for a set of quantum numbers that distinguish independent
eigenfunctions of the lowest order Schr\"odinger equation, and a
subscript $i$ of $E^k_i$ and $\psi^k_i$ for the order of $1/m_Q$.
\subsection{-1st order}
\label{subsec:-1st}
{}From eqs.(\ref{Schr}) and (\ref{ham_1}),
the -1st order Schr\"odinger equation in $1/m_Q$ is given by
\begin{equation}
    -2m_Q\Lambda_-\psi^k_0=0,
\end{equation}
which means
\begin{equation}
    \psi^k_0=\Lambda_+\psi^k_0.
\end{equation}
That is, the 0-th order wave function has only a positive component of
the heavy quark sector whose form is given below.
\subsection{0-th order}
\label{subsec:0-th}
The 0-th order equation is given by
\begin{equation}
    -2m_Q\Lambda _-\psi _1^k+H_0\psi _0^k=E_0^k\psi _0^k.
\end{equation}
Multiplying projection operators, $\Lambda_\pm$, from left, respectively,
we obtain
\begin{eqnarray}
  &&\Lambda _+H_0\psi _0^k=E_0^k\psi _0^k,\label{0th:1}\\
  &&-2m_Q\Lambda _-\psi _1^k+\Lambda _-H_0\psi _0^k=0.\label{0th:2}
\end{eqnarray}
Eq.\ (\ref{0th:1}) gives the lowest non-trivial Schr\"odinger equation.
Let us assume this equation is solved\cite{Matsuki1} and all the independent
eigenfunctions and eigenvalues are obtained analytically and/or
numerically as $E^k_0$ and
\begin{equation}
    \psi _0^k=\Phi^+_k=\left( {\matrix{{u_k\left({\vec r,\Omega }\right)}&0\cr
    {v_k\left( {\vec r,\Omega } \right)}&0\cr }} \right),
\end{equation}
where $u_k$ and $v_k$ are 2 by 2 matrices.
Likewise we can in general assume that a $\Lambda_-$ component of
any wave function is expanded in terms of
\begin{equation}
    \Phi^-_k=\left( {\matrix{0&{u_k\left( {\vec r,\Omega } \right)}\cr
    0&{v_k\left( {\vec r,\Omega } \right)}\cr}} \right).
\end{equation}
Expanding $\Lambda_-\psi _1^k$ in terms of this set of eigenfunctions,
one can solve Eq.\ (\ref{0th:2}) as follows. Setting
\begin{equation}
    \Lambda _-\psi _1^k=\sum\limits_\ell  {c_{1-}^{\ell \,k}\Phi _\ell ^-},
\end{equation}
one obtains coefficients, $c^{\ell\,k}_1$, as
\begin{equation}
    c_{1-}^{\ell \,k}={1 \over {2m_Q}}\int {r^2drd\Omega }
    \;{\rm tr}\left({\Phi  ^- _\ell}^\dagger\Lambda _-H_0\Lambda_+\Phi^+_k
    \right),
\end{equation}
where the 0-th order wave function is normalized to be 1,
\begin{equation}
     \int d^3 r~{\rm tr}
     \left({\Phi_k^\pm\,}^\dagger\Phi_\ell^\pm\right)
     =\delta_{k\,\ell}.\label{nomal0}
\end{equation}
\subsection{1st order}
\label{subsec:1st}
The 1st order equation is given by
\begin{equation}
    -2m_Q\Lambda _-\psi _2^k+H_0\psi _1^k+H_1\psi _0^k
    =E_0^k\psi _1^k+E_1^k\psi _0^k.
\end{equation}
As in the above case, multiplying projection operators,
we obtain
\begin{eqnarray}
  &&\Lambda _+H_0\psi _1^k+\Lambda _+H_1\psi _0^k
  =E_0^k\Lambda _+\psi _1^k+E_1^k\psi _0^k,\label{1st:1}\\
  &&-2m_Q\Lambda _-\psi _2^k+\Lambda _-H_0\psi _1^k+\Lambda _-H_1\psi _0^k
  =E_0^k\Lambda _-\psi _1^k.\label{1st:2}
\end{eqnarray}
When one looks at Eq.\ (\ref{1st:1}), one easily notices that
this can be separated into two independent equations.
One equation gives a solution to the positive component of $\psi^k_1$ as
\[    \Lambda _+H_0\Lambda _+\psi _1^k=E_0^k\Lambda _+\psi _1^k,\]
which means from Eq.\ (\ref{0th:1}) that
\begin{equation}
    \Lambda _+\psi _1^k=c^k_{1+} \Phi^+_k,
\end{equation}
and the coefficient, $c^k_{1+}$, should be determined by a normalization
of the wave function up to the first order,
\begin{equation}
    \int d^3r~{\rm tr}\left({\psi^k}^\dagger\psi^k\right)=1,\label{normal}
\end{equation}
whose definition is allowed because here we are not calculating the
absolute value of
the form factors. The appropriate normalization will be determined
in future papers in which we will give several kinds of form factors.
Using this equation, one easily notices that up to the first order,
\begin{equation}
    c^k_{1+}=0,
\end{equation}
where the coefficient is assumed to be real.
This completes the solution for $\psi _1^k$ since $\Lambda_-\psi^k_1$
is obtained in the last chapter, i.e.,
\begin{equation}
    \psi _1^k=\sum\limits_\ell
    {c_{1-}^{\ell \,k}\Phi_\ell ^-}.
\end{equation}
Another equation is given by
\begin{equation}
    \Lambda _+H_0\Lambda _-\psi _1^k+\Lambda _+H_1\psi _0^k=E_1^k\psi _0^k,
    \label{pert:1st}
\end{equation}
which gives the first order perturbation corrections to the mass when
one calculates matrix elements of the lhs among eigenfunctions,
$\Phi^\pm_k$ like in the ordinary perturbation of quantum mechanics.

Eq.\ (\ref{1st:2}) gives a $\Lambda_-$ component of $\psi^k_2$
as in the case of $\Lambda_-\psi^k_1$ in subsection \ref{subsec:0-th}, i.e.,
setting
\begin{equation}
    \Lambda _-\psi _2^k=\sum\limits_\ell  {c_{2-}^{\ell \,k}\Phi _\ell ^-},
\end{equation}
one obtains coefficients, $c^{\ell\,k}_{2-}$, as
\begin{equation}
    c_{2-}^{\ell \,k}={1 \over {2m_Q}}\int {r^2drd\Omega }
    \;{\rm tr}\left[{\Phi  ^- _\ell}^\dagger \left(\Lambda _-H_0\psi^k_1
    +\Lambda _-H_1\Phi^+_k-E^k_0\Lambda _-\psi^k_1\right)\right].
\end{equation}
\subsection{2nd order}
\label{subsec:2nd}
The 2nd order equation is given by
\begin{equation}
    -2m_Q\Lambda _-\psi _3^k+H_0\psi _2^k+H_1\psi _1^k+H_2\psi _0^k
    =E_0^k\psi _2^k+E_1^k\psi _1^k+E_2^k\psi_0^k.
\end{equation}
As in the above cases, multiplying projection operators, we obtain
\begin{eqnarray}
  &&\Lambda _+H_0\psi _2^k+\Lambda _+H_1\psi _1^k+\Lambda_+H_2\psi _0^k
  =E_0^k\Lambda _+\psi _2^k+E_1^k\Lambda _+\psi _1^k
  +E_2^k\psi_0^k,\label{2nd:1}\\
  &&-2m_Q\Lambda _-\psi _3^k+\Lambda _-H_0\psi _2^k+\Lambda _-H_1\psi _1^k
  +\Lambda _-H_2\psi _0^k=E_0^k\Lambda _-\psi _2^k
  +E_1^k\Lambda _-\psi _1^k.\label{2nd:2}
\end{eqnarray}
When one looks at Eq.\ (\ref{2nd:1}), one easily notices that this
can be separated into three independent equations.
The first one gives a solution to the positive component of $\psi^k_2$:
\[    \Lambda _+H_0\Lambda _+\psi _2^k=E_0^k\Lambda _+\psi _2^k,\]
which means from Eq.\ (\ref{0th:1}) that
\begin{equation}
    \Lambda _+\psi _2^k=c^k_{2+} \Phi^+_k.
\end{equation}
The coefficient, $c^k_{2+}$, should be determined by a normalization
of the wave function up to the second order using Eq.\ (\ref{normal})
and by assuming the coeffcient is real, it is given by
\begin{equation}
    c^k_{2+} = - {1\over 2}\left(\sum\limits_\ell c_{1-}^{\ell \,k}\right)^2.
\end{equation}
There are some other contributions to this coefficient, $c^k_{2+}$,
from transitions of $\Phi_k^+$ to other states and back to $\Phi_k^+$
which are obtained when one diagonalizes the whole energy matrix.
This completes the solution for $\psi _2^k$
when combined with the $\Lambda_-$ component obtained above.
\begin{equation}
  \psi^k_2=c^k_{2+} \Phi^+_k +
  \sum\limits_\ell  {c_{2-}^{\ell \,k}\Phi _\ell ^-}.
\end{equation}
The second equation is given by
\[  a\Lambda _+H_1\Lambda _-\psi _1^k+\Lambda _+H_1\Lambda _+\psi _1^k
  =E_1^k\Lambda _+\psi _1^k,\]
where $a$ is determined so that this equation coincides with
the first order energy correction equation, Eq.\ (\ref{pert:1st}),
using $\Lambda_+\psi^k_1=c^k_{1+}\psi^k_0$:
\[  a=c_{1+}^k=0.\]
The third equation is given by the remainder,
\begin{equation}
    \Lambda _+H_0\Lambda _-\psi _2^k+
    \Lambda _+H_1\Lambda _-\psi _1^k+\Lambda_+H_2\psi _0^k
    =E_2^k\Lambda _+\psi _0^k,\label{pert:2nd}
\end{equation}
which gives the second order corrections to the energy by taking the
matrix elements of the lhs among $\Phi^\pm_k$.

Eq.\ (\ref{2nd:2}) gives a $\Lambda_-$ component of $\psi^k_3$
as in the cases of $\Lambda_-\psi^k_1$ and $\Lambda_-\psi^k_2$
given in subsections \ref{subsec:0-th} and \ref{subsec:1st}, i.e.,
setting
\begin{equation}
    \Lambda _-\psi _3^k=\sum\limits_\ell  {c_{3-}^{\ell \,k}\Phi _\ell ^-},
\end{equation}
one obtains coefficients, $c^{\ell\,k}_{3-}$, as
\begin{equation}
    c_{3-}^{\ell \,k}={1 \over {2m_Q}}\int {r^2drd\Omega }
    \;{\rm tr}\left[{\Phi  ^- _\ell}^\dagger \left(\Lambda _-H_0\psi^k_2
    +\Lambda_-H_1\psi^k_1
    +\Lambda _-H_2\Phi^+_k-E^k_0\Lambda _-\psi^k_2
    -E^k_1\Lambda _-\psi^k_1\right)\right].
\end{equation}
\subsection{Higher order}
\label{subsec:higher}
The $i$-th order equation is given by
\begin{equation}
    -2m_Q\Lambda _-\psi ^k_{i+1}+\sum\limits_{j=0}^i {H_j\psi ^k_{i-j}}
    =\sum\limits_{j=0}^i {E^k_j\psi ^k_{i-j}}.
\end{equation}
We follow the previous subsections to derive all the results from this
equation below. Multiplying projection operators, we obtain
\begin{eqnarray}
  &&\sum\limits_{j=0}^i {\Lambda _+H_j\psi ^k_{i-j}}
  =\sum\limits_{j=0}^i {E^k_j\Lambda _+\psi ^k_{i-j}},\label{higher:1}\\
  &&-2m_Q\Lambda _-\psi _{i+1}^k+\sum\limits_{j=0}^i {\Lambda
_-H_j\psi^k_{i-j}}
  =\sum\limits_{j=0}^{i-1} {E^k_j\Lambda _-\psi ^k_{i-j}}.\label{higher:2}
\end{eqnarray}
Eq.\ (\ref{higher:1}) gives a solution to a positive component of $\psi^k_i$ as
\begin{equation}
    \Lambda _+H_0\Lambda _+\psi _i^k=E_0^k\Lambda _+\psi _i^k,
\end{equation}
which means
\begin{equation}
    \Lambda _+\psi _i^k=c^k_{i+} \Phi^+_k,
\end{equation}
where the coefficient, $c^k_{i+}$, should be determined by a normalization
of the wave function, Eq.\ (\ref{normal}), up to the $i$-th order and
there are of course other contributions mentioned in subsection
\ref{subsec:2nd}. This completes the solution for $\psi _i^k$
when combined with the $\Lambda_-$ component obtained from the $(i-1)$-th order
equation. Eq.\ (\ref{higher:1}) also includes other equations,
which are identified as the $(i-j)$-th order $(j=1\sim i-1)$
energy correction equations. These are straightforwardly
obtained order by order though complicated and hence are omitted here.
After extracting these equations, the remainder of Eq.\ (\ref{higher:1})
gives the $i$-th order energy correction equation.
Eq.\ (\ref{higher:2}) gives a $\Lambda_-$ component of $\psi^k_{i+1}$, i.e.,
setting
\begin{equation}
    \Lambda _-\psi _{i+1}^k=\sum\limits_\ell  {c_{i+1,-}^{\ell \,k}
    \Phi _\ell ^-},
\end{equation}
one obtains coefficients, $c^{\ell\,k}_{i+1,-}$, as
\begin{equation}
    c_{i+1,-}^{\ell \,k}={1 \over {2m_Q}}\int {r^2drd\Omega }
    \;{\rm tr}\left[{\Phi  ^- _\ell}^\dagger \left(\sum\limits_{j=0}^{j=i-1}
    \Lambda _-\left(H_j-E^k_j \right)\psi^k_{i-j}
    +\Lambda _-H_i\Phi^+_k\right)\right],
\end{equation}
where all the $E_j^k$ are the values obtained by diagonalizing the lhs of
the $j$-th order energy correction equation.
\section{Conclusions}
\label{sec:Comments}
In this letter we have developed a systematic perturbation of the
Shcr\"odinger equation for a bound state of $Q\bar q$ at rest in $1/m_Q$
by using just once the naive Foldy-Wouthuysen-Tani transformation on
the heavy quark sector. The effects of the FWT transformation is,
as can be seen from Eq.\ (\ref{hameqs}), to treat the heavy quark as a
non-relativistic color source. It is shown that all the positive components of
wave functions at each order of perturbation are proportional to the
lowest order solution, $\Phi^+_k$ except that $c^k_{1+}=0$.
There also appear negative components of wave functions for heavy quark
sector, which can also be perturbatively solved. As one can easily notices,
the lowest order equation, Eq.\ (\ref{0th:1}), can be derived by any
method. That is, we do not need to introduce the FWT or any other
transformation to obtain this equation. The higher order terms in $1/m_Q$,
however, reflect what kind of transformation is introduced and we believe that
any transformation which utilizes the properties of a heavy quark will
give a better perturbation as has been shown by HQET than
that without a transformation.

To make formulae simple in this letter, all the gamma matrices of
light and heavy quarks are multiplied from left with the wave function,
$\psi^k$. To make the wave function a true bispinor,
the Schr\"odinger equation should be transformed by a charge
conjugation operator $U_c=i \gamma^0_Q\gamma^2_Q$ for a heavy quark.
That is, $U_c\,\psi^k$
becomes a bispinor, $U_c\,H_{FWT}\,U_c^\dagger$ is a hamiltonian of interest,
and gamma matrices of a light anti-quark is multiplied from left
while those of a heavy quark from right.

One example for model dependent interaction terms is, e.g.,
given by\cite{Morii1,Morii2}
\begin{equation}
    \beta_q\beta_Q~S(r) + \left\{1-{1\over
2}\left[\vec\alpha_q\cdot\vec\alpha_Q
    +(\vec\alpha_q\cdot\vec n) (\vec\alpha_Q\cdot\vec
    n)\right]\right\}~V(r),
\end{equation}
where $\vec n=\vec r/r$.
This model is being studied by using our formulation, in which
masses and eigenfunctions of $D$, $D^*$, $B$ and $B^*$
are calculated including all those corrections
mentioned in Sec.\ref{sec:pert} and spin-flavor symmetry at
the lowest order is realized by a special quantum number related to the
heavy quark spin.\cite{Matsuki1} Our formulation gives
the same results as those in \cite{Morii1} if negative components
are neglected and if only the terms up to $1/m_Q^2$ in the hamiltonian
are taken into account.

In order to calculate, e.g., the semileptonic weak decay $B\to D+\ell\nu$,
one needs to boost the $D$ meson even if the $B$ meson is at rest.
The formulation obtained in this letter does not give any prescription
how to construct a boosted wave function. We will
propose some method in a future paper to compute the form factors
by using wave functions calculated by the method proposed in this letter.
\acknowledgments
The author would like to thank K. Akama and T. Morii
for helpful discussions and also the theory group
at Institute for Nuclear Study for a warm hospitality where
a part of this work was done.
%
% references
%


\begin{references}
\bibitem{HQET}
	M. Voloshin and M. Shifman, {\sl Sov. J. Nucl. Phys.} {\bf 45},
        292 (1987) and {\bf 47}, 511 (1988);\\
	N. Isgur and M. Wise, {\sl Phys. Lett.} {\bf B232}, 113 (1989)
	and {\bf B237}, 527 (1990);\\
	E. Eichten and B. Hill, {\sl Phys. Lett.} {\bf B234}, 511 (1990);\\
	B. Grinstein, {\sl Nucl. Phys.} {\bf B339}, 253 (1990);\\
	H. Georgi, {\sl Phys. Lett.} {\bf B240}, 447 (1990);\\
	A. Falk, H. Georgi, B. Grinstein and M. Wise, {\sl Nucl. Phys.}
        {\bf B343}, 1 (1990).
\bibitem{POL}
	M. E. Luke, {\sl Phys. Lett.} {\bf B252}, 447 (1990);\\
	J. G. K\"orner and G. Thompson, {\sl Phys. Lett.}
        {\bf B264}, 185 (1991);\\
	T. Mannel, W. Roberts and Z. Ryzak, {\sl Nucl. Phys.} {\bf 368},
        204 (1992);\\
	S. Balk, J. G. K\"orner and D. Pirjol,
	{\sl Nucl. Phys.} {\bf B428}, 499 (1994);
\bibitem{REVIEW} See for review,\\
	N. Isgur and M. B. Wise, {\sl B Decays}, (ed. S. Stone,
        World Scientific, Singapore, 1992);\\
        H. Georgi, {\sl Perspective in the Standard Model},
        (ed. R. K. Ellis, C. T. Hill and J. D. Lykken,
        World Scientific, Singapore, 1992).
\bibitem{FWT}
	L. L. Foldy and S. A. Wouthuysen, {\sl Phys. Rev.} {
        \bf 78}, 29 (1950);\\
	S. Tani, {\sl Prof. Theor.} Phys. {\bf 6}, 267 (1951).
\bibitem{Morii1}
    J. Morishita, M. Kawaguchi and T. Morii, {\sl Phys. Rev.}
    {\bf 37}, 159 (1988).
\bibitem{Matsuki1}
	A talk given by T. Matsuki at {\sl Confinement'95} and also at
        {\sl Wein'95} held at RCNP (Osaka Univ.), TKU-95-1 preprint, Sept.
1995,
	to be published in the {\sl Proceedings of Wein'95};\\
	T. Matsuki and T. Morii, TKU-95-3 preprint in preparation.
\bibitem{Morii2}
    S. N. Mukherjee, R. Nag, S. Sanyal, T. Morii, J. Morishita and M. Tsuge,
    {\sl Phys. Rep.} {\bf 231}, 201 (1993).
\end{references}
\end{document}